\documentclass[]{spie}  
\usepackage[]{graphicx}
\usepackage{subfig,booktabs,ctable,multirow,amsmath,newclude}

\DeclareMathOperator*{\argmax}{arg\,max}
\DeclareMathOperator*{\argmin}{arg\,min}

\title{PCA/LDA Approach for Text-Independent Speaker Recognition}

\author{Zhenhao Ge, Sudhendu R. Sharma, Mark J.T. Smith
\skiplinehalf
School of Electrical and Computer Engineering \\
Purdue University, West Lafayette, IN, USA, 47907
}

\authorinfo{Further author information: Zhenhao Ge: zge@purdue.edu, Sudhendu R. Sharma: sharmasr@purdue.edu}

\begin{document}
\maketitle 

\begin{abstract}

Various algorithms for text-independent speaker recognition have been developed through the decades, aiming to improve both accuracy and efficiency. This paper presents a novel PCA/LDA-based approach that is faster than traditional statistical model-based methods and achieves competitive results. First, the performance based on only PCA and only LDA is measured; then a mixed model, taking advantages of both methods, is introduced. A subset of the TIMIT corpus composed of 200 male speakers, is used for enrollment, validation and testing. The best results achieve $100\%,96\%$ and $95\%$ classification rate at population level $50,100$ and $200$, using 39-dimensional MFCC features with delta and double delta. These results are based on 12-second text-independent speech for training and 4-second data for test. These are comparable to the conventional MFCC-GMM methods, but require significantly less time to train and operate.

\end{abstract}



\section{Introduction}
\label{sec:introduction} 

Speaker recognition is a growing research area with applications in access control, transaction authentication, law enforcement, speech data management, to mention a few\cite{reynolds2001automatic}. Speaker recognition typically encompasses \textit{speaker classification} and \textit{speaker verification}\cite{campbell1997speaker}. The former one classifies the testing speaker into one of the pre-modelled classes, or identify the testing speaker as a new speaker (in the open-set case\footnote{open-set case is to decide whether or not the unknown testing speaker belong to a set of $S$ known speakers, while the close-set case is to classify the input speaker to one of the speakers pre-modelled in the classifier.}),  while the latter one makes a binary decision on whether the input speaker is the the speaker he/she claims to be. Speaker verification can be considered to be a special case of speaker classification in an open-set case. Speaker recognition is often sub-divided into text-independent and text-dependent cases, based on whether or not the speech used is known for each speaker. This paper mainly focuses on text-independent speaker classification in a close-set case, which means the input speaker must be pre-modelled and included in the classifier.

Many statistical model approaches have been considered in this area, such as these based on Gaussian Mixture Models (GMMs)\cite{reynolds1995robust}, Hidden Markov Models (HMMs)\cite{savic1990variable,benzeghiba2006speechcom}, Support Vector Machines (SVMs)\cite{wan2000support}, Artificial Neural Networks (ANN)\cite{farrell1994speaker}, and so on. These algorithms generally achieve high accuracy but usually require a significant amount of time to train and test. For some cases like meetings and conferences, there may be a need for very fast speaker recognition response times in which case a more efficient and low cost approach may be desirable.

There have been extensive studies that demonstrate that both Principle Component Analysis (PCA) and Linear Discriminant Analysis (LDA) are helpful in improving the efficiency of speaker recognition system while maintaining high accuracy. Zhang et al.\cite{zhang2003exploiting} investigated a PCA-based classifier involving both a Principle Component Space (PCS) and a Truncation Error Space (TES), and showed that this mixed classifier can outperform either of the two individual PCS or TES classifier. Jin et al.\cite{jin2000application} reported that the GMM classifier with LDA feature reduction can achieve higher performance with respect to accuracy and efficiency in some circumstances. Other researchers have integrated PCA with GMM\cite{seo2001gmm} and genetic algorithms\cite{islam2010noise}, and have applied PCA and LDA in conjunction with K-Nearest Neighbors (KNN) algorithms\cite{kacur2011speaker} for speaker identification.

\begin{figure}[!h]
 \centering
   \includegraphics[scale=0.8]{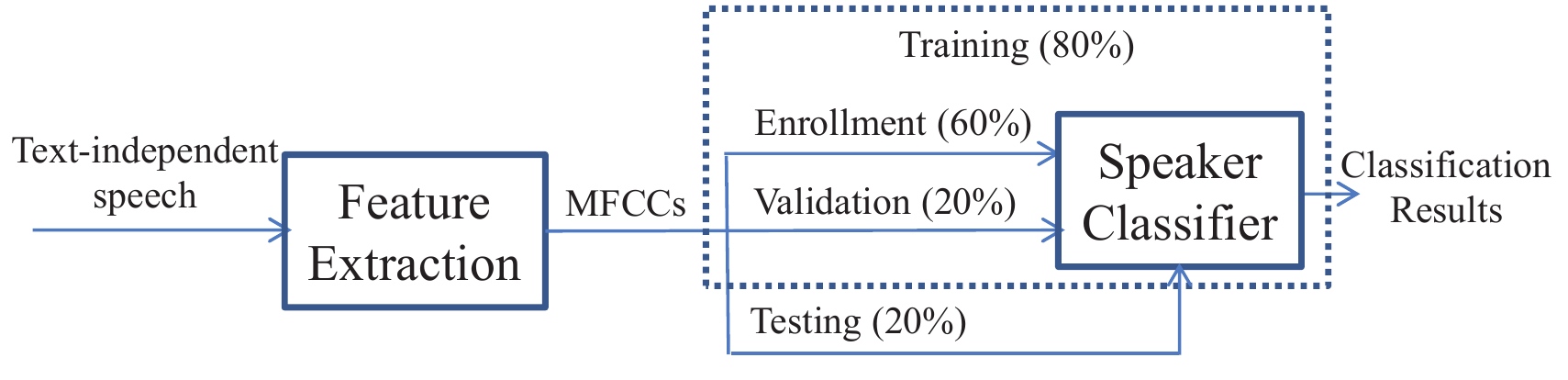}
   \caption{General structure used for text-independent speaker classification \label{fig:structure}}		
\end{figure}

This paper extends some of this previous work in new directions. We first explore text-independent speaker classification using PCA or LDA individually with optimized parameter settings. Then, we will combine these two classifiers into one and demonstrate that the composite achieves comparable results to conventional GMM approaches while significantly reducing computation time. The general structure of the speaker classification system used in this paper is shown in Fig.~\ref{fig:structure}. We used $80\%$ of the data from each speaker to train the PCA-based, LDA-based and combined classifiers, $60\%$ for enrollment and $20\%$ for validation. The remaining  $20\%$ is used to test the performance of the final composite classifier. 

\section{PCA Method for Text-Independent Speaker Recognition}
\label{sec:pca} 

In this section, we introduce a PCA classification method and demonstrate how to compute PCA eigenspaces and transform data to the eigenspace with reduced dimension. Then, a PCA classifier based on both Principle Component Space (PCS) and Truncation Error Space (TES) is explored and evaluated.

\subsection{Introduction of PCA}
\label{subsec:pca_introduction}

Principle Component Analysis (PCA) is a standard data analysis technique that transforms the original data set $\mathcal{D}\{\mathbf{\Gamma}_1,\mathbf{\Gamma}_2,...\mathbf{\Gamma}_N\}$ to an $K$-dimensional uncorrelated and orthogonal space spanned by $U_K$ using eigenvalue decomposition. The dimension of the transformed data set $\mathcal{D'}\{\mathbf{\omega}_1,\mathbf{\omega}_2,...,\mathbf{\omega}_N\}$ is normally reduced and sorted by the corresponding variance, i.e., the eigenvalue of the data covariance matrix $C$, on that dimension. A new eigenspace $U_k$ with reduced  dimension $k$ is formed by the $k$ normalized eigenvector of $C$ associated with the $k$ largest eigenvalues.

This process of data transformation or projection from the original space to the  eigenspace can be implemented as follows:

\begin{enumerate}\itemsep0pt
\item Assuming each original sample $\mathbf{\Gamma}_i$, $i \in [1,N]$ is $M$-dimensional, find the sample mean $\mathbf{\Psi} = \frac{1}{M}\sum^M_{i=1}\mathbf{\Gamma}_i$;
\item Shift $\mathbf{\Gamma}_i$ to zero mean and form a new data matrix $A = [\mathbf{\Phi}_i,\mathbf{\Phi}_2,...,\mathbf{\Phi}_N]$ ($M \times N$), where $\mathbf{\Phi}_i = \mathbf{\Gamma}_i - \mathbf{\Psi}$;
\item Find the covariance matrix $C = \frac{1}{M}\sum^M_{i=1}\mathbf{\Phi}_i\mathbf{\Phi}^T_i \propto AA^T$ (ignore the constant factor $\frac{1}{M}$);
\item Use an eigen-decomposition method, such as Singular Value Decomposition (SVD), to find the eigenvalues $\Lambda = [\lambda_1,\lambda_2,...,\lambda_K]$, $K = \min(M,N)$,  and the corresponding eigenvectors $U_K = [\mathbf{u}_1,\mathbf{u}_2,...,\mathbf{u}_K]$ ($M \times K$) of $C$ in descending order w.r.t $\Lambda$;
\item Choose the first $k (k \leq K)$ eigenvectors to form the eigenspace $U_k = [\mathbf{u}_1,\mathbf{u}_2,...,\mathbf{u}_k]$ ($M \times k$); 
\item Project the mean-shifted data set $A$ ($M \times N$) to the eigenspace $U_k$ ($M \times k$) and represent the data by $\Omega = U_k^TA$ ($k \times N$) or $\mathbf{\omega}_i = U_k^T\mathbf{\Phi}_i$,$i \in [1,N]$.
\end{enumerate}

Data components on the dimensions with larger eigenvalues contain significant information about the data. Removing components on the dimensions with smaller eigenvalues maintains principle information while reducing dimension. With the assumption that the principle information also helps to distinguish classes, one can:

\begin{enumerate}\itemsep0pt
\item Compute an overall eigenspace with data from all classes and classify the transformed data based on the distance from eigenspace ($e_\mathrm{dfes}$) and the distance within eigenspace ($e_\mathrm{dies}$), using some pattern recognition algorithm, such as K-Nearest Neighbors (KNN), or Gaussian Mixture Models (GMMs);
\item Compute an individual eigenspace for each classes $s$ and form a PCA classifier $g^{(s)}(X)$ to perform classification directly by computing $\hat{S} = \mathrm{arg\,max}\,g^{(s)}(X)$, in which $s \in [1,S]$, $S$ is the total number of classes, and $X$ is the input feature set.
\end{enumerate}


The first approach is usually applied when the  data set of all classes share similar characteristics that can be used to distinguish them from random data using $e_\mathrm{dfes}$, e.g., face images are easily distinguished from random image in face recognition. In addition, data from the same class are normally clustered in the eigenspace computed by all classes. For example, an input face image can be determined by measuring its distance $e_\mathrm{dies}$ to each centroid of the face class in the eigenspace.

However, for data such as Mel-Frequency Cepstral Coefficients (MFCCs) used in text-independent speaker recognition, the major variation of data is the content of speech  rather than the characteristics of the speakers, and the projected data samples from different classes are significantly overlapped in the overall eigenspace. Thus, we can only choose the 2nd approach to decorrelate and orthogonalize data using PCA, explore which dimensions contribute more than the others, and finally form a PCA-based classifier to reduce the dimensionality of data and maintain good speaker recognition performance at the same time. 

\subsection{PCA-based Classifier}
\label{subsec:pca_explore}

As mentioned in Section \ref{subsec:pca_introduction}, the PCA-based classifier requires the computation of eigenspace $U_K(s)$ for each class $s$, $s \in [1,S]$, i.e., for each speaker. Twelve seconds of text-independent speech for each speaker $s$ is converted to an $M \times N$ MFCCs feature matrix with delta and double delta using a 10 msec frame rate, where $M$ and $N$ are the dimension and length (number of frames) of the cepstral features. After mean-shifting, covariance matrix computation,  and eigenvalue decomposition, the full eigenspace $U^{(s)}$ for each speaker $s$ can be found and a input sample $\mathbf{\Gamma}_i$ can be classified by measuring its ``distance" of mean-shifted features $\mathbf{\Phi}_i$ to each eigenspace $U^{(s)}$.

\begin{figure}[!h]
 \centering
   \includegraphics[scale=0.7]{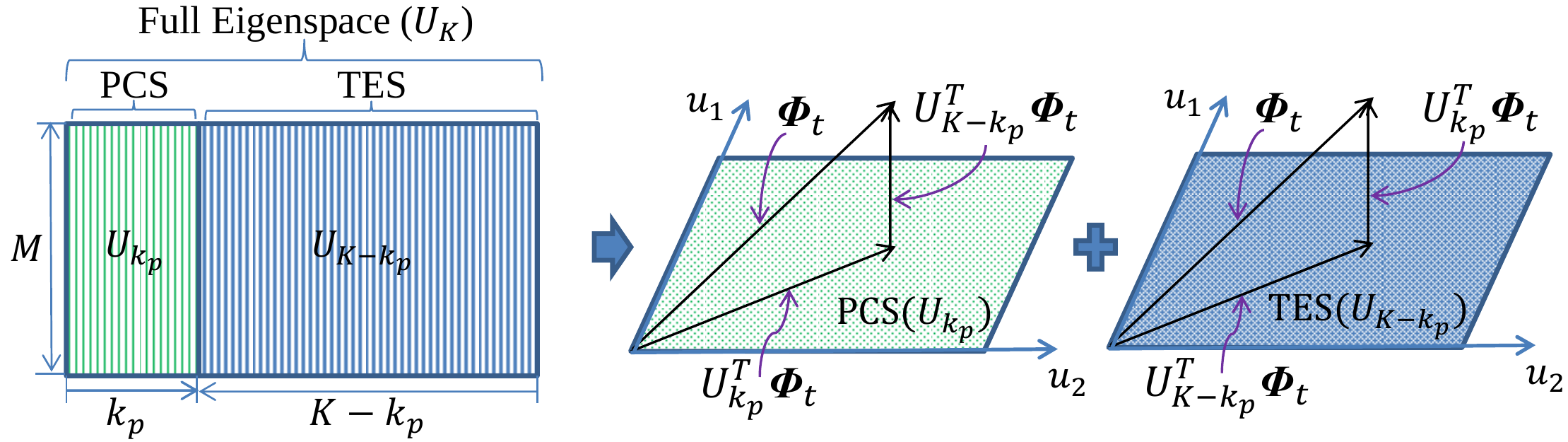}
   \caption{Projections onto the $k_p$-dimensional Principle Component Space (PCS) and the ($K-k_p$)-dimensional Truncation Error Space (TES) \label{fig:pcs-tes_1}}		
\end{figure}

Once the full eigenspace $U_K$ is reduced to the Principle Component Space (PCS) with dimension $k_p$, the rest of the $K-k_p$ eigenvectors form another eigenspace called the Truncation Error Space (TES), which is orthogonal to the PCS. After the samples $\mathbf{\Gamma}$ from one input speaker being converted to $X$, two intuitive classification approaches are considered: maximize the projection of $X$ onto PCS; and  minimize its projection onto the TES. The relationships among the PCS, the TCS and the  full eigenspace, and the projections onto these spaces are illustrated in Fig.~\ref{fig:pcs-tes_1}. 

The PCS and TES classifers may be described as:
\begin{equation} \label{eq:pcs}
	\textrm{PCS Classifier:} \; \hat{S} 
	= \argmax_{s \in [1,S]} \, g^{(s)}_{_{\mathrm{PCS}}}(X), \;
	\mathrm{where} \; g^{(s)}_{_{\mathrm{PCS}}}(X) 
	= \sum^T_{t=1} \Vert U^{(s)}_{k_p}\mathbf{\Phi}^{(s)}_t \Vert 
	= \sum^T_{t=1} \Vert U^{(s)}_{k_p}(\mathbf{x}_t-\mathbf{\Psi}^{(s)}) \Vert,
\end{equation}
\begin{equation} \label{eq:tes}
	\textrm{TES Classifier:} \; \hat{S} 
	= \argmin_{s \in [1,S]} \, g^{(s)}_{_{\mathrm{TES}}}(X), \;
	\mathrm{where} \; g^{(s)}_{_{\mathrm{TES}}}(X) 
	= \sum^T_{t=1} \Vert U^{(s)}_{K-k_p}\mathbf{\Phi}^{(s)}_t \Vert 
	= \sum^T_{t=1} \Vert U^{(s)}_{K-k_p}(\mathbf{x}_t-\mathbf{\Psi}^{(s)}) \Vert.
\end{equation}
In Eq.~(\ref{eq:pcs}) and Eq.~(\ref{eq:tes}), $U^{(s)}_{k_p}$ and $U^{(s)}_{K-k_p}$ denote the PCS and TES of speaker $s$, $\mathbf{x}_t$ and $\mathbf{\Phi}_t$ denote the $t$th feature vector of $X$ ($M \times T$) before and after mean-shift based on the speaker $s$.

\begin{figure}[!h]
 \centering
   \includegraphics[scale=0.7]{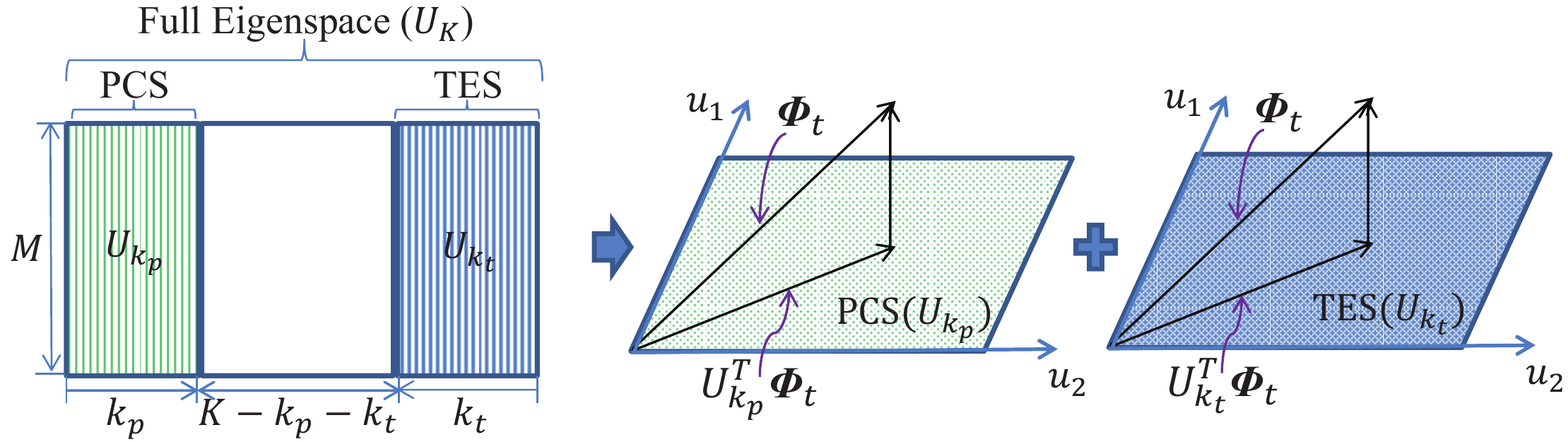}
   \caption{Projections onto $k_p$-dimensional PCS and $k_t$-dimensional TES \label{fig:pcs-tes_2}}		
\end{figure}

Zhang, et al. \cite{zhang2003exploiting} have shown combining these two criteria can reach higher classification performance than either of the two individual classifiers based on PCS or TES. Here we further investigate how each dimension component in PCS and TES contributes to classification accuracy. We propose a new mixed classifier defined by the equation
\begin{equation} \label{eq:pcs-tes}
	\textrm{PCS\&TES Classifier:} \; \hat{S} 
	= \argmax_{s \in [1,S]} \, g^{(s)}(X|\lambda)
	= \argmax_{s \in [1,S]} \, p \, g^{(s)}_{_{\mathrm{PCS}}}(X|k_p) - 
	(1-p) \, g^{(s)}_{_{\mathrm{TES}}}(X|k_t), \;
	\mathrm{where} \; \lambda = \{k_p,k_t,p\}.
\end{equation}
In Eq.~(\ref{eq:pcs-tes}), parameter set $\lambda$ contains $k_p$, $k_t$ and $p$, which denote the dimensions of PCS and TES and the weight of these two individual classifiers. The relation between $k_p$ and $k_t$ and the projection of $\mathbf{\Phi}_t$ onto both eigenspaces spanned by $U_{k_p}$ and $U_{k_t}$ is illustrated in Fig.~\ref{fig:pcs-tes_2}. Next, we intend to find $\lambda^*$ which optimizes the mixed classifier and renders best performance, through exhausted search in the 3-dimensional space of ($k_p$, $k_t$, $p$).

\begin{figure}[!ht]
  \centering
  \subfloat[population size: 100; 39 MFCCs ($\Delta^2$)]{
  \label{fig:kpkt_s100}\includegraphics[width=0.5\textwidth]{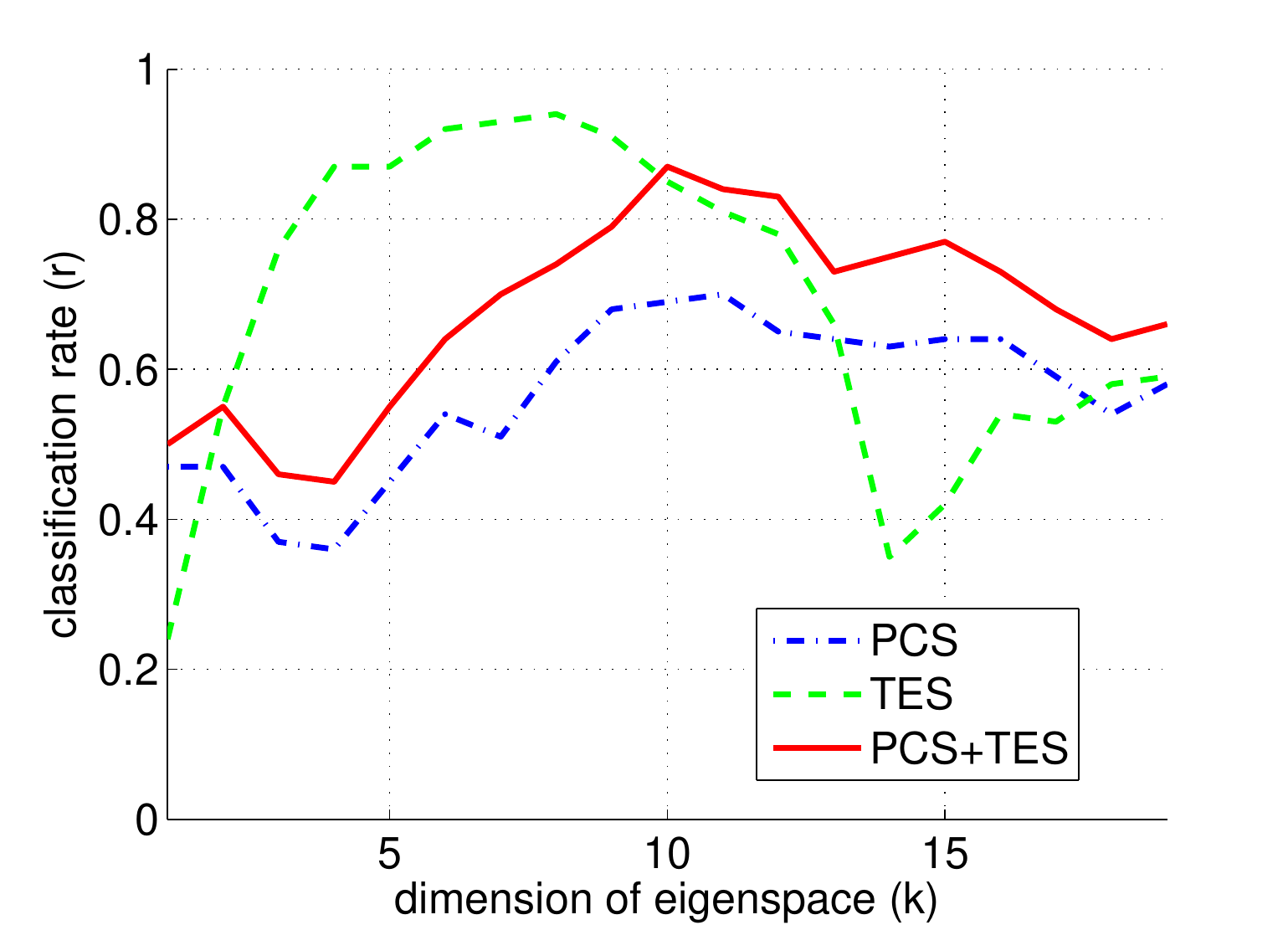}}                    
  \subfloat[population size: 200; 39 MFCCs ($\Delta^2$)]{
  \label{fig:kpkt_s200}\includegraphics[width=0.5\textwidth]{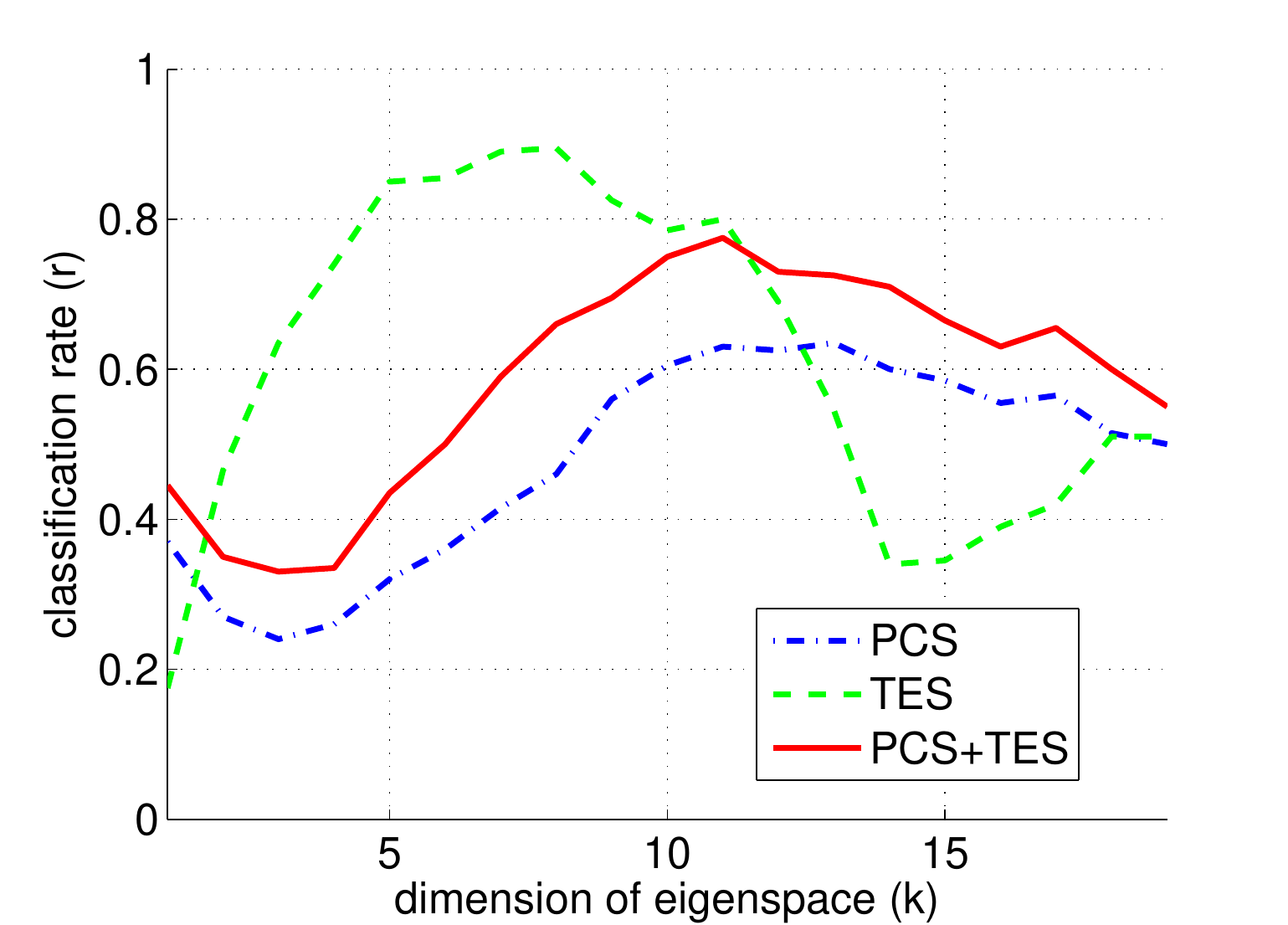}}
  \caption{Performance of PCS and TES classifiers w.r.t. dimension}
  \label{fig:kpkt}
\end{figure}

In order to find appropriate search ranges for $k_p$ and $k_t$, the performance in terms of classification rate ($r$) of PCS and TES individually along with different dimensions is investigated is illustrated in Fig.~\ref{fig:kpkt}. Three curves of classification rates are presented, including PCS (blue dash-dot line), TES (green dashed line) and the combination of PCS and TES with $p = 0.5$ (red solid line). It indicates that the combined classifier may outperform the single classifier at certain dimensions, and also shows both PCS and TES classifiers reach peak performance within dimension range (1,$\lfloor M/2 \rfloor$), where $M = 39$ is the feature dimension used in this project. Thus, it is sufficient to implement exhausted search in $k_p, k_t \in (1,\lfloor M/2 \rfloor$) and $p \in [0,1]$ with interval $0.01$.

\subsection{Performance Evaluation of the PCA-based classifier}
\label{subsec:pca_performance}

Classification performances using PCS and TES individually, and the mixed model w.r.t parameter set $\lambda = \{k_p,k_t,p\}$ are compared and listed below in Table~\ref{tab:pca_comp1}. Based on the results, the mixed classifier is in general better than the other two, while the TES classifier is better than the PCS, and thus contributes more in the mixed classifier. The dimension $k_t$ in TES is relatively consistent compared with the dimension in PCS as the population size increases. For the mixed classifier at each population level, there are multiple points of $(k_p,k_t,p)$ that reach the top performance, w.r.t. both accuracy in terms of classification rate and efficiency in terms of number of total dimension $k_p+k_t$. A more detailed list of all points of ($k_p,k_t,p$) is included in Appendix~\ref{sec:a1_pca_comp2}. 

\begin{table}[!h]
\centering
\caption{Performance comparison for PCS,TES and mixed classifiers with optimized parameter settings}
\label{tab:pca_comp1}
\begin{tabular}{@{} c|cc|cc|ccccc @{}} \toprule%
\multicolumn{1}{>{\bfseries}c}{Pop. Size} & 
\multicolumn{1}{>{\bfseries}c}{\%  $ r_{_{\mathrm{PCS}}}$} & 
\multicolumn{1}{>{\bfseries}c}{$k_p$} & 
\multicolumn{1}{>{\bfseries}c}{\% $r_{_{\mathrm{TES}}}$} & 
\multicolumn{1}{>{\bfseries}c}{$k_t$} & 
\multicolumn{1}{>{\bfseries}c}{\% $r_{_{\mathrm{Mixed}}}$} & 
\multicolumn{1}{>{\bfseries}c}{$k_p+k_t$} & 
\multicolumn{1}{>{\bfseries}c}{$k_p$} & 
\multicolumn{1}{>{\bfseries}c}{$k_t$} & 
\multicolumn{1}{>{\bfseries}c}{\% $p$} \\\midrule 
50 & 82.0 & 9 & 96.0 & 8 & \textbf{100} & 9 & 1 & 8 & 37 $\leftrightarrow$ 76 \\
100 & 70.0 & 11 & 94.0 & 8 & \textbf{99.0} & 13 & 5 & 8 & 2 $\leftrightarrow$ 6 \\
200 & 63.5 & 13 & 89.5 & 8 & \textbf{96.5} & 17 & 9 & 8 & 14 $\leftrightarrow$ 15
\\\bottomrule
\end{tabular}
\end{table}

\section{LDA Method for Text-Independent Speaker Recognition}
\label{sec:lda}

While PCA seeks a dimension-reduced orthogonal eigenspace with largest data variance in each direction; the goal of Linear Discriminant Analysis (LDA) is to find another $K$-dimensional eigenspace ($K \leq M$, the data dimension), with the first $K$ directions that maximally discriminate among different classes. In this section, we first introduce LDA, then project MFCCs to a new eigenspace based on LDA with fewer dimensions. After that, we use a Gaussian Mixture Model (GMM)-based classifier with dimension reduced features to perform speaker classification.  

\subsection{Introduction of LDA}
\label{subsec:lda_introduction}

For this discussion of  Linear Discriminant Analysis (LDA), assume there is $M \times T_s$ data $X$ for class $s \in [1,S]$, where $M$ is the sample dimension and $T_s$ is the number of samples in this class $s$. $\Phi$ and $\Phi_s$ are the global mean over all classes and the local mean for each class $s$ respectively. Then, we define between-class scatter $S_B$ and within-class scatter $S_W$ by 
\begin{equation} \label{eq:bcs}
	S_B = \frac{1}{S}\sum^S_{s=1}(\Phi_s-\Phi)(\Phi_s-\Phi)^T, 
\end{equation}
\begin{equation} \label{eq:wcs}
	S_W = \frac{1}{S}\sum^S_{s=1}\frac{1}{T_s}\sum^{T_s}_{t=1}(X_t-\Phi_s)(X_t-\Phi_s)^T.
\end{equation}
If we choose $\mathbf{w}$ from the underlying space $W$, then $\mathbf{w}^T S_B \mathbf{w}$ and $\mathbf{w}^T S_W \mathbf{w}$ are the projections of $S_B$ and $S_W$ onto the direction $\mathbf{w}$. Searching the directions $\mathbf{w}$ for the best class discrimination is equivalent to maximizing the ratio of $(\mathbf{w}^T S_B \mathbf{w}) / ({\mathbf{w}^T S_W \mathbf{w}})$ subject to $\mathbf{w}^T S_W \mathbf{w} = 1$. The latter is called the Fisher Discriminant Function and can be converted to 
\begin{equation} \label{eq:lda}
	S_B\mathbf{w} = \lambda S_W \mathbf{w}, \;\mathrm{then}\;
	S^{-1}_W S_B \mathbf{w} = \lambda \mathbf{w} 
\end{equation}
by Lagrange multipliers and solved by eigen-decomposition of $S^{-1}_W S_B$.

Using this process on the $M$-dimensional MFCCs feature set of speaker data, we find the eigenspace $W_K$ ($K \leq M$) and reduce the feature dimension from $M$ to $K$ by projecting them to the eigenspace. Then, we use Gaussian Mixture Models (GMMs), which have been successfully used to classify speakers based on MFCCs \cite{reynolds1995robust} to perform speaker recognition. Since the construction of eigenspaces in LDA requires information from all classes, we cannot construct LDA eigenspaces for each speaker and form a LDA-based classifier using the method described for the PCA-based classifier. Instead, we use LDA for dimension reduction  prior to GMM-based classification. 

\subsection{GMM for speaker recognition}
\label{subsec:gmm4spkr}

Gaussian mixture density models the feature distribution of each speaker as a weighted sum of multiple Gaussian distributions. For each feature vector $\mathbf{x}$ in the $M \times T$ feature set $X$, the probability of $\mathbf{x}$ can be formulated by the equation 
\begin{equation} \label{eq:gmm}
	p(\mathrm{x}|\lambda) = \sum^N_{i=1} p_i b_i(\mathbf{x}),\;
	b_i(\mathbf{x}) = \frac{1}{(2\pi)^{M/2}\vert\Sigma_i\vert^{1/2}} \mathrm{exp}\lbrace - \frac{1}{2}(\mathbf{x}-\mathbf{\mu}_i)^T\Sigma^{-1}_i(\mathbf{x}-\mathbf{\mu}_i)\rbrace,
\end{equation}
where $M$ is the dimension of the feature vector $\mathbf{x}$, $N$ is the number of mixture components, $b_i(\mathbf{x})$, $i = 1,...,N$, are the component densities, $p_i$, $i=1,...,N$, are the mixture weights and $\lambda = \{p_i,\mu_i,\Sigma_i$\}, $i = 1,2,...,N$ is the collective representation of the parameters.

GMMs are attractive for modelling speakers because they may reveal the underlying vocal tract configurations, which help to distinguish speakers; and they are capable  of representing a large class of sample distributions\cite{reynolds1995robust}.  

Given MFCCs feature $X$ ($M \times T$) from speaker $s$, the Maximum Likelihood Estimation (MLE) is used to maximize the GMM likelihood, which can be written as 
\begin{equation} \label{eq:mle}
	\lambda^* = \argmax_\lambda p(\mathbf{X}|\lambda) 
	= \argmax_\lambda \prod^T_{t=1}p(\mathbf{x}_t|\lambda).
\end{equation}
Since this expression is non-linear and direct maximization is difficult, the parameter set $\lambda = \{p,\mu,\Sigma\}$ is iteratively estimated using a special case of the Expectation-Maximization (EM) algorithm \cite{dempster1977maximum} and is summarized below:  
\begin{equation} \label{eq:gmm}
	\bar{p}_i = \frac{1}{T}\sum^T_{t=1}p(i|\mathbf{x}_t,\lambda); \;
	\bar{\mathbf{\mu}}_i = \frac{\sum^T_{t=1}p(i|\mathbf{x}_t,\lambda)\mathbf{x}_t}{\sum^T_{t=1}p(i|\mathbf{x}_t),\lambda)}; \;
	\bar{\mathbf{\sigma}}^2_i = \frac{\sum^T_{t=1}p(i|\mathbf{x}_t,\lambda)\mathbf{x}^2_t}{\sum^T_{t=1}p(i|\mathbf{x}_t,\lambda)}-\bar{\mathbf{\mu}}^2_i,
\end{equation}
where $\bar{p}_i,\bar{\mathbf{\mu}}_i,\bar{\mathbf{\sigma}}^2_i$, $i = 1,...,N$ are the mixture weights, means, and variances for the $i$th component. $p(i|\mathbf{x}_t,\lambda)$ is \textit{a posteriori} probability for the $i$-th component given by 
\begin{equation} \label{eq:posteriori}
	p(i|\mathbf{x}_t,\lambda) = \frac{p_ib_i(\mathbf{x}_t)}{\sum^M_{k=1}p_kb_k(\mathbf{x}_t)}\;.
\end{equation}
These estimates are based on the assumption of independence among feature dimension, so for each speaker class $s$, the non-zero values of the covariance matrix are only on the diagonals. 

This algorithm guarantee a monotonic increase of the model's likelihood on each EM iteration. Detailed implementation with parameter initialization is discussed in Sec.~\ref{sec:implementation}. 

After obtaining the GMM parameter set $\lambda_s$ for speaker class $s \in [1,S]$, the GMM-based classifier, which maximize \textit{a posteriori} probability for a feature sequence $X$, ($M \times T$) can be formulated as follows:
\begin{equation} \label{eq:gmmclassifier}
	\textrm{GMM Classifier:} \;
	\hat{S} = \argmax_{s \in [1,S]}\mathrm{Pr}(\lambda_s|X)
			= \argmax_{s \in [1,S]}\frac{p(X|\lambda_s)\mathrm{\lambda_s}}{p(X)}
			\propto \argmax_{s \in [1,S]}p(X|\lambda_s)
			\propto \argmax_{s \in [1,S]}\sum^T_{t=1}\mathrm{log}p(\mathbf{x}_t|\lambda_s).
\end{equation}
The first equation is due to Bayes' rule; the first proportion is assuming $\mathrm{Pr}(\lambda_s) = 1/S$ and $p(X)$ is the same for all speaker models; the second proportion uses logarithm and independence between input samples $\mathbf{x}_t$, $t \in [1,T]$.

The LDA-GMM classifier is similar to the GMM classifier in Eq.~(\ref{eq:gmmclassifier}). The only difference is the input features $X$ are replaced by feature $Y$ with fewer dimensions reduced by LDA.

\begin{figure}[!ht]
  \centering
  \includegraphics[scale=0.7]{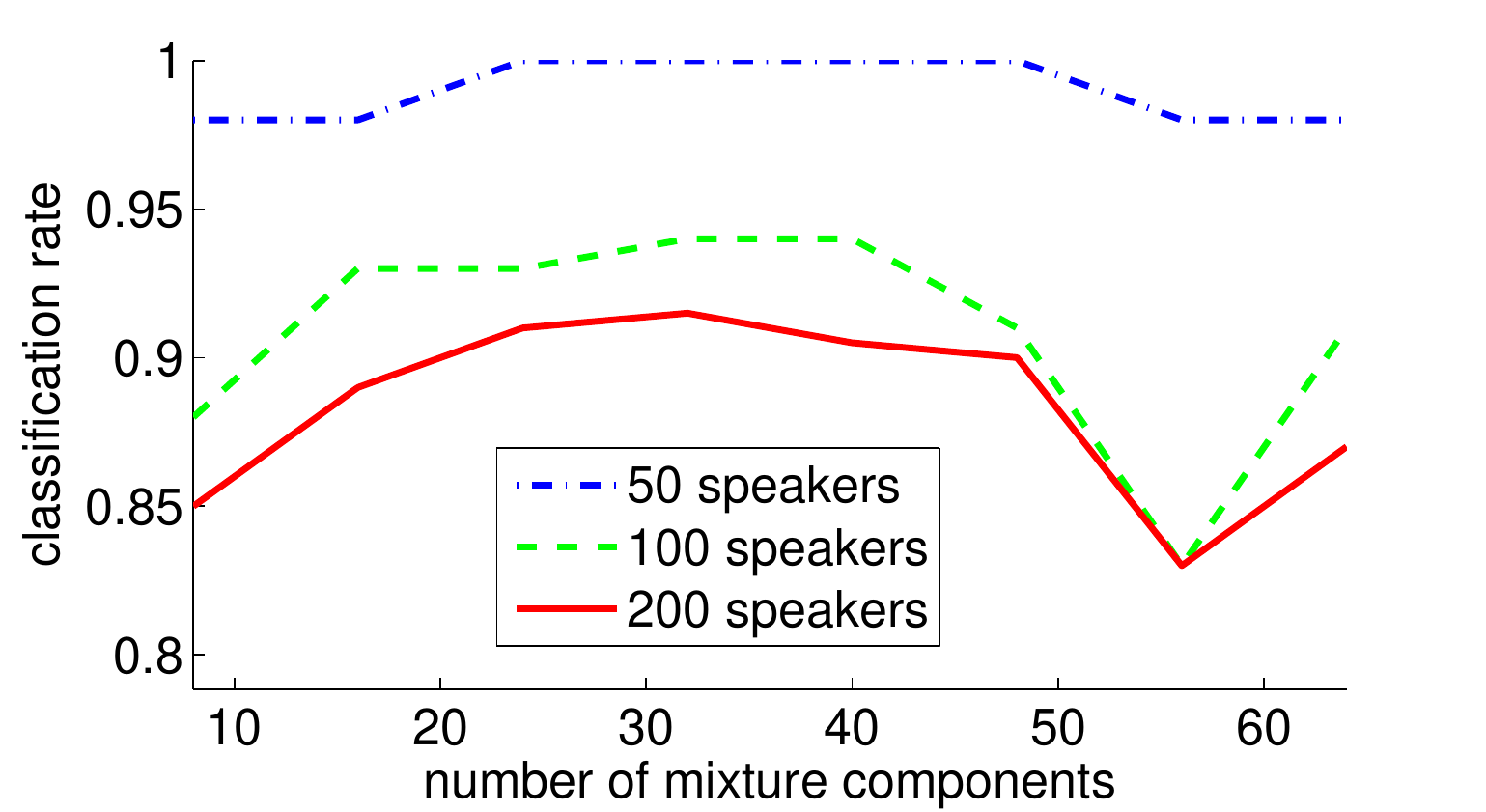}
   \caption{Performance vs. number of GMM mixture components for three population sizes}
  \label{fig:selectM}
\end{figure}

One of the important issues when using a GMM as a classifier is determining the approximate number of mixtures $N$, i.e., the model order. Applying a GMM on the original feature data without LDA dimension reduction, and plotting performance vs. model order ($N = 8,16,...,64$) curves in multiple population levels (Fig. \ref{fig:selectM}), we found the range of $N \in [15,40]$ is appropriate for our database. Using LDA prior to the GMM yields similar results. Thus, we select the GMM with model order $N = 15$ and $N = 30$ to evaluate LDA-GMM classifier in the next section. 

\subsection{Performance Evaluation of the LDA-GMM classifier}
\label{subsec:lda-gmm_performance}

\begin{figure}
  \centering
  \subfloat[number of mixtures: 15; 39 MFCCs ($\Delta^2$)]{
  \label{fig:rk_m15}\includegraphics[width=0.5\textwidth]{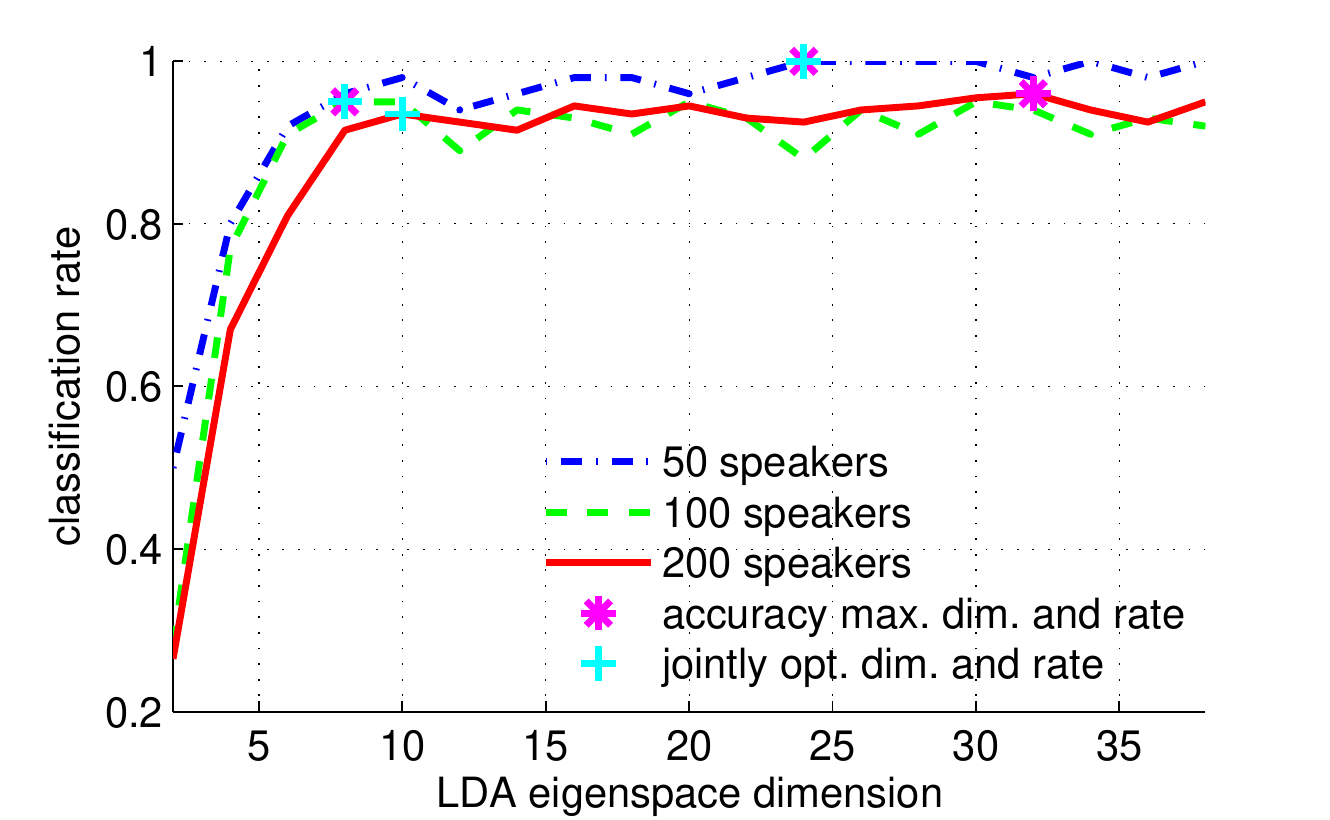}}                    
  \subfloat[number of mixtures: 30; 39 MFCCs ($\Delta^2$)]{
  \label{fig:fig:rk_m30}\includegraphics[width=0.5\textwidth]{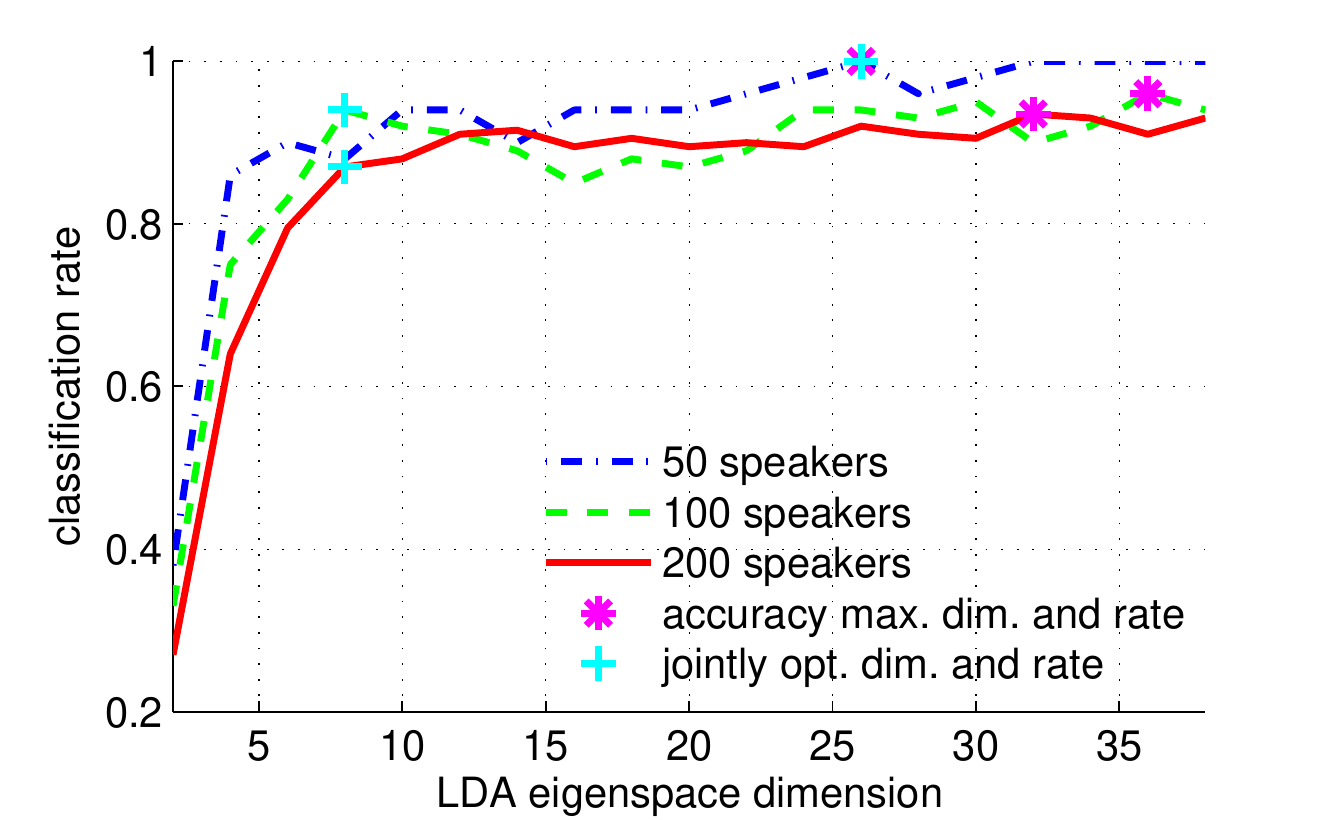}}
  \caption{Performance v.s. LDA eigenspace dimension with two GMM model orders}
  \label{fig:rk}
\end{figure}

Jin et al. have shown that applying LDA can not only reduce the feature dimension and the computation cost, but also improve the classification rate\cite{jin2000application}. In this paper, we aim to quantitatively find the ``optimal" reduced feature dimension $k$ ($k<M$), that balances both classification accuracy and efficiency. Fig.~\ref{fig:rk} shows the classification rate $r$ of the LDA-GMM classifier increases along with the LDA eigenspace dimension $k$. 
If the dimension is optimized by maximizing the accuracy, it may be close to the original full dimension $M$ and the computational cost is similar to the GMM classifier without LDA. Thus, a joint performance/complexity optimized  dimension $k$ is developed by minimizing the product of the error rate ($1-r$) and dimension $k$ and is formulated below:
\begin{equation} \label{eq:kopt}
	k^* = \argmin_{k \in [1,M]}(1-r)k.
\end{equation}
The ``plus signs" in Fig.~\ref{fig:rk} show the jointly optimized dimension and the corresponding classification rate $r$, while the ``stars" show the solutions with $r$ maximized.

In Table~\ref{tab:lda_gmm_performance}, the performance of the LDA-GMM classifier with optimized LDA dimensions using both criteria is compared with the GMM classifier.
\begin{table}
\begin{small}
\begin{center}
\caption{Performance comprison of the LDA-GMM and the GMM classifiers with three population sizes}
\label{tab:lda_gmm_performance}
\begin{tabular}{@{} *{7}{c} @{}} \toprule
Pop. Size ($S$) & \multicolumn{2}{c}{50} & \multicolumn{2}{c}{100} & \multicolumn{2}{c}{200} \\
Model Order ($N$) & 15 & 30 & 15 & 30 & 15 & 30 \\\midrule
$\% r_{_{\mathrm{GMM}}} [k]$ & 100 [39] & 98 [39] & 96 [39] & 96 [39] & 94 [39] & 93 [39] \\
$\% r_{_{\mathrm{LDA-GMM}}}(\textrm{accuracy max.}) [k]$ & 100 [24] & 100 [26] & 95 [8] & 96 [36] & 96 [32] & 93.5 [32] \\
$\% r_{_{\mathrm{LDA-GMM}}}(\textrm{jointly opt.}) [k]$ & 100 [24] & 100 [26] & \textbf{95 [8]} & 94 [8] & \textbf{93.5 [10]} & 87 [8] \\\bottomrule 
\end{tabular}
\end{center}
\end{small}
\end{table} 
When model order $N=30$, the LDA-GMM classifier with maximized accuracy always performs better than or equal to the GMM classifier, while when $M=15$, it only performs better than the GMM classifier at speaker level $S = 200$. This indicates that applying LDA is more useful when the model order or population size is large. In terms of efficiency, by comparing $r_{_{\mathrm{GMM}}}$ and $r_{_{\mathrm{LDA-GMM}}} (\textrm{jointly opt.})$ at population size 100 and 200, reducing feature dimension when the model order is low will maintain competitive accuracy with good efficiency. The LDA-GMM classifier configuration highlighted digits will be used to implement the mixed PCA/LDA approach discussed in Sec.~\ref{sec:implementation}.

\section{Implementation of the PCA/LDA Combined Method}
\label{sec:implementation} 

Though PCA and LDA are commonly used for feature dimension reduction, both of them have their own advantages and disadvantages and it has been a long debate over decades on which one is better\cite{o2007theoretical,yang2003can}. PCA is relatively easy to implement, since the matrix used in eigen-decomposition is always non-singular. This is not the case for LDA. Moreover, PCA requires less computation, especially when we compute PCA eigenspace for each class and form a PCA-based classifier without other pattern classification techniques, such as the PCA classifier designed in this paper. However, the discriminant information may not reside in the direction with large component variance. That is the weakness of PCA and where LDA shines. LDA suffers from the issue of singularity and the ``peaking" problem \cite{sima2008peaking}, when the training database is small. Recent studies show PCA may outperform LDA when the training set is small and PCA is less sensitive to different training sets\cite{martinez2001pca}.   

Since PCA and LDA are two complementary techniques, after discussing text-independent speaker recognition based on each of them in Sec.~\ref{sec:pca} and Sec.~\ref{sec:lda}, we present the detailed experimental setting to implement PCA/LDA combined method with better results than single PCA and LDA classifiers.

\subsection{Database and Feature Extraction}

A subset of the TIMIT corpus consisting of 200 Male speakers with 10 utterances each from region 1 to 4 is used for implementation. The 10 utterances from each speaker are then divided into 3 parts, one for enrollment, one for validation and one for testing. 

\begin{enumerate}\itemsep0pt
\item The first 12 seconds from the concatenation of No. 3 to No. 8 utterances is used for enrollment, such as computing PCA, LDA eigenspace, or training the GMM models for each speaker.
\item The first 4 seconds from the concatenation of No.1 and No. 2 utterances is used for validation, such as computing the classification rate of the PCA and LDA-GMM classifiers and determining the settings of parameters $\lambda_{_{\mathrm{PCA}}}=\{k_p,k_t,p\}$ and  $\lambda_{_{\mathrm{LDA-GMM}}}=\{N,k\}$ and the weight $p$ in the PCA/LDA combined classifier.
\item The first 4 seconds from the concatenation of No. 9 and No. 10 utterances is used for testing the performance of the PCA/LDA combined classifier, which is illustrated later in this section.
\end{enumerate}

Thirty-nine dimensional HTK-style MFCCs with delta and double delta at 10 msec frame rate are used \cite{ellis@mfcc}. The cepstral coefficients from order 2 to 13 (removing order 1 of the DC component) plus 1 energy feature comprise the original 13-dimensional MFCCs. 13-dimensional delta (or velocity) feature and another 13-dimensional double delta (or acceleration) feature are added in to measure the changes between frames in the corresponding cepstral/energy ferature\cite{jurafsky2008speech}. The Hamming window length is 25 msec. So for each speaker, the sizes of the feature sets for enrollment, validation and testing are $39 \times 1198$, $39 \times 398$ and $39 \times 398$.

\subsection{GMM Initialization and Training}

Given an $M \times T$ feature set $X$ for each speaker, where $M$ is the dimension of each feature vector and $T$ is the number of vectors (samples), to initialize an $N$-dimensional GMM, $N$ evenly spaced feature vectors $\mathbf{x}_i$, $i \in [1,N]$ from $X$ are selected to be the initial mean $\mathbf{\mu}_i$, The initial variance matrix $\Sigma_i$ is $M \times M$ identity matrix, and the weight $p_i$ is $1/N$. 

In GMM training, a variance limiting constraint $\mathrm{Var}(\mathbf{x})/T^2$ is used on each dimension of $X$ to avoid singularity in the model's likelihood function. The maximum number of iterations in the EM algorithm to find the MLE of the GMM is 100 with an early termination condition that the increment of log-likelihood between two consecutive iterations is less than 0.001. 

\subsection{Combined Classifier based on PCA/LDA}

\begin{figure}
  \centering
  \includegraphics[scale=0.9]{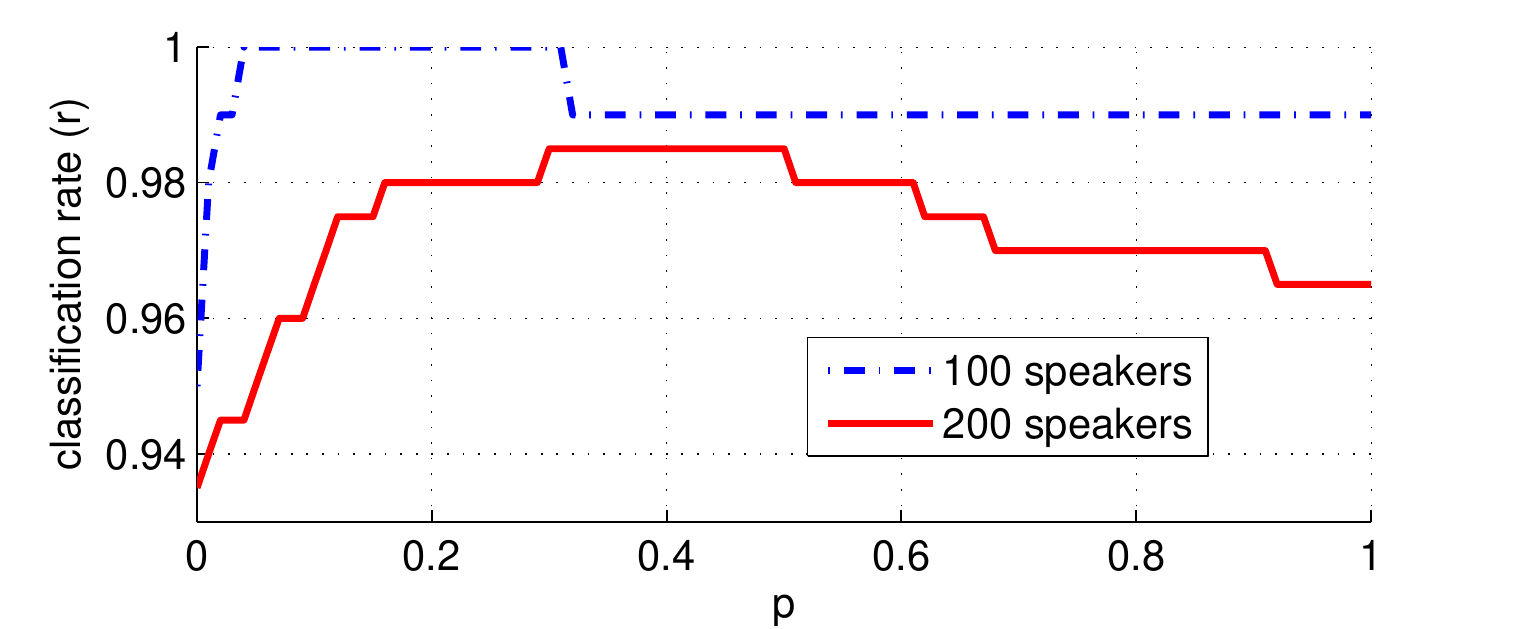}
   \caption{Performance of PCA/LDA combined classifier w.r.t. the weight $p$}.
  \label{fig:pca-lda_performance}
\end{figure}

To combine the PCA-based (PCS\&TES) classifier and the LDA-GMM-based classifier, we first normalized them to uniform scale using the following equation:
\begin{equation} \label{eq:norm}
	g^{(s)}_1 = \frac{g^{(s)}_{_{\mathrm{PCS\&TES}}}(X)}{\sum^S_{s=1}(g^{(s)}_{_{\mathrm{PCS\&TES}}}(X))^2}, \;
	g^{(s)}_2 = \frac{g^{(s)}_{_{\mathrm{LDA-GMM}}}(X)}{\sum^S_{s=1}(g^{(s)}_{_{\mathrm{LDA-GMM}}}(X))^2},
\end{equation}
where $g_1$ and $g_2$ are the normalized classifiers of both approaches. Then, we optimized the weight $p$ on the combined classifier by
\begin{equation} \label{eq:norm}
	p^* = \argmax_{p \in [0,1]}\,r(p)
	    = \argmax_{p \in [0,1]}\,\frac{\sum^S_{s=1}I_{s=\argmax_{s \in [1,S]}pg^{(s)}_1(X)+(1-p)g^{(s)}_2(X)}}{S},
\end{equation}
where $r(p)$ is the classification rate of combined classifier based on $p$ and $I_{s=\argmax_{s \in [1,S]}pg^{(s)}_1(X)+(1-p)g^{(s)}_2(X)}$ is an indicator that speaker $s$ is correctly classified using the combined classifier. 
Fig.~\ref{fig:pca-lda_performance} shows the ranges of $p^*$ with 100 and 200 population sizes are $4\%$ to $31\%$ and $30\%$ to $50\%$ respectively, given the parameter set $\{k_p,k_t,p|S\}$ for $g_1$ is $\{5,8,4\%|S=100\}$ and $\{9,8,15\%|S=200\}$, and the parameter set $\{N,k|S\}$ for $g_2$ is $\{15,8|S=100\}$ and $\{15,10|S=200\}$.

For population 100 and 200, we select $p^* = 17.5\%$ and $p^* = 40\%$, which are the medians in the range of the optimized $p$, as the final $p^*$ in the combined classifier. The PCA/LDA combined classifier is given by:
\begin{equation} \label{eq:pca-lda}
	\textrm{PCA/LDA Combined Classifier:} \;
	\hat{S} = \argmax_{s \in [1,S]}\,p^*g^{(s)}_1(X) + (1-p^*)g^{(s)}_2(X).
\end{equation}
When tested using the remaining two utterances for each speaker, the combined classifier achieves $100\%,96\%,95\%$ classification rate for population size of $50,100,200$, which is slightly better than GMM classifier ($100\%,96\%,94\%$ given model order $N=15$, refer to Table~\ref{tab:lda_gmm_performance}), but with significantly less computation time, about $80\%$ less than the GMM classifier in this implementation.

\section{Conclusion and Future Work}
\label{sec:conclusion} 

This paper presents classifiers based on PCA and LDA with optimized parameter settings. In the PCA approach, a classifier based on both PCS and TES is discussed and evaluated with globally optimized settings in $k_p$,$k_t$ the dimension of PCS and TES and the weight $p$, which balances the two individual PCS and TES classifiers. In the LDA approach, the feature dimension is reduced using LDA, before being passed on as an input to the GMM classifier. The GMMs are initialized with appropriate model order with consideration of accuracy and efficiency. Then, a LDA-GMM classifier with optimized settings is shown to achieve comparable accuracy in speaker classification with significantly reduced time. After the development of both PCA and LDA-GMM classifier, a classifier combining these two techniques achieve higher performance w.r.t. both accuracy and efficiency. Using only 12 seconds of  text-independent utterances for training, and 4 seconds for testing, the combined classifier achieves $100\%,96\%$ and $95\%$ classification rates for population sizes of $50,100$ and $200$, which is comparable to the performance of the conventional GMM classifier on the same data. However, the new combined system reduces the computation by up to $85\%$ in training and $78\%$ in testing, compared with GMM classifier.

For future work, there are at least three types of evaluation on the combined classifier  worth exploring, which may eventually provide a more robust PCA/LDA-based speaker classifier for various databases:
\begin{enumerate}\itemsep0pt
\item One can experiment with different lengths of speech in both training and testing and obtain an optimized balance between the accuracy and efficiency for specific databases;
\item One can also investigate the classifier with different types of features, such as Linear Predictive Coding Coefficients (LPCCs) and Perceptual Linear Prediction (PLP), even using MFCCs, one can explore how the delta and double delta features can help to distinguish classes;
\item In terms of robustness, the classifier should also be tested with multiple databases with different noise levels.
\end{enumerate}

\appendix  

\section{Parameter Setting of PCA-based Classifier with Optimized Performance}
\label{sec:a1_pca_comp2}

When implementing exhausted search to find the optimized parameter set $\lambda = \{k_p,k_t,p\}$ that provides the best classification performance, multiple points in the parameter space were found. Some of them were clustered together, showing a stable region that reaches high performance, while the other were scattered. Here we provide a more detailed list of these points of parameter setting than shown in Table~\ref{tab:pca_comp1} in Sec.~\ref{subsec:pca_performance}. These points have been sorted in ascending order of computational cost w.r.t. $k_p+k_t$. After considering both classification accuracy and efficiency, we have selected some points to form a combined classifier with LDA and the results are provided in Sec.~\ref{sec:implementation}.

\begin{table}[!ht]
\begin{small}
\begin{center}
\caption{List of PCA classifier parameters with best performance}
\label{tab:pca_comp2}
\begin{tabular}{@{} *{5}{c}|*{5}{c}|*{5}{c} @{}} \toprule
\multicolumn{5}{l}{Pop. Size: 50} & \multicolumn{5}{l}{Pop. Size: 100} & \multicolumn{5}{l}{Pop. Size: 200} \\
\multicolumn{5}{l}{\% Best Performance: 100} & \multicolumn{5}{l}{\% Best Performance: 99} & \multicolumn{5}{l}{\% Best Performance: 96.5} \\
\multicolumn{5}{l}{Total points: 446 (70 shown)} & \multicolumn{5}{l}{Total points: 28} & \multicolumn{5}{l}{Total points: 5} \\\midrule
$k_p+k_t$ & $kp$ & $k_t$ & $\% p$ & pts & $k_p+k_t$ & $kp$ & $k_t$ & $\% p$ & pts & $k_p+k_t$ & $kp$ & $k_t$ & $\% p$ & pts \\
9 & 1 & 8 & 37 $\leftrightarrow$ 76 & 40 & 7 & 1 & 6 & 49 & 1 & 17 & 9 & 8 & 14 $\leftrightarrow$ 15 & 2 \\
12 & 4 & 8 & 5 $\leftrightarrow$ 9 & 5 & 13 & 5 & 8 & 2 $\leftrightarrow$ 6 & 5 & 18 & 10 & 8 & 17 $\leftrightarrow$ 18  & 2 \\
13 & 5 & 8 & 2 $\leftrightarrow$ 11 & 10 & 14 & 6 & 8 & 4 $\leftrightarrow$ 6 & 3 & 19 & 11 & 8 & 12 & 1 \\
14 & 6 & 8 & 3 $\leftrightarrow$ 10 & 8 & 16 & 8 & 8 & 3 $\leftrightarrow$ 4 & 2 & & & & & \\
15 & 7 & 8& 4 $\leftrightarrow$ 10 & 7 & 17 & 9 & 8 & 5 $\leftrightarrow$ 6 & 2 & & & & & \\
& & & & & 26 & 18 & 8 & 5 $\leftrightarrow$ 9 & 15 \\\bottomrule
\end{tabular}
\end{center}
\end{small}
\end{table}

\bibliography{reference}

\begin{thebibliography}{10}

\bibitem{reynolds2001automatic}
Reynolds, D., ``Automatic speaker recognition: Current approaches and future
  trends,'' {\em Speaker Verification: From Research to Reality}  (2001).

\bibitem{campbell1997speaker}
Campbell~Jr, J., ``Speaker recognition: A tutorial,'' {\em Proceedings of the
  IEEE}~{\bf 85}(9),  1437--1462 (1997).

\bibitem{reynolds1995robust}
Reynolds, D. and Rose, R., ``Robust text-independent speaker identification
  using gaussian mixture speaker models,'' {\em Speech and Audio Processing,
  IEEE Transactions on}~{\bf 3}(1),  72--83 (1995).

\bibitem{savic1990variable}
Savic, M. and Gupta, S., ``Variable parameter speaker verification system based
  on hidden markov modeling,'' in [{\em Acoustics, Speech, and Signal
  Processing, 1990. ICASSP-90., 1990 International Conference
  on}{\nolinebreak\hspace{0.1em}]},   281--284, IEEE (1990).

\bibitem{benzeghiba2006speechcom}
BenZeghiba, M.~F. and Bourlard, H., ``User-customized password speaker
  verification using multiple reference and background models,'' {\em Speech
  Communication}~{\bf 8} (0 2006).
\newblock IDIAP-RR 04-41.

\bibitem{wan2000support}
Wan, V. and Campbell, W., ``Support vector machines for speaker verification
  and identification,'' in [{\em Neural Networks for Signal Processing X, 2000.
  Proceedings of the 2000 IEEE Signal Processing Society
  Workshop}{\nolinebreak\hspace{0.1em}]},   {\bf 2},  775--784, IEEE (2000).

\bibitem{farrell1994speaker}
Farrell, K., Mammone, R., and Assaleh, K., ``Speaker recognition using neural
  networks and conventional classifiers,'' {\em Speech and Audio Processing,
  IEEE Transactions on}~{\bf 2}(1),  194--205 (1994).

\bibitem{zhang2003exploiting}
Zhang, W., Yang, Y., and Wu, Z., ``Exploiting pca classifiers to speaker
  recognition,'' in [{\em Neural Networks, 2003. Proceedings of the
  International Joint Conference on}{\nolinebreak\hspace{0.1em}]},   {\bf 1},
  820--823, IEEE (2003).

\bibitem{jin2000application}
Jin, Q. and Waibel, A., ``Application of lda to speaker recognition,'' in [{\em
  Sixth International Conference on Spoken Language
  Processing}{\nolinebreak\hspace{0.1em}]},  (2000).

\bibitem{seo2001gmm}
Seo, C., Lee, K., and Lee, J., ``Gmm based on local pca for speaker
  identification,'' {\em Electronics Letters}~{\bf 37}(24),  1486--1488 (2001).

\bibitem{islam2010noise}
Islam, R. and Rahman, F., ``Noise robust speaker identification using pca based
  genetic algorithm,'' {\em International Journal of Computer Applications
  IJCA}~{\bf 4}(12),  27--31 (2010).

\bibitem{kacur2011speaker}
Kacur, J., Vargic, R., and Mulinka, P., ``Speaker identification by k-nearest
  neighbors: Application of pca and lda prior to knn,'' in [{\em Systems,
  Signals and Image Processing (IWSSIP), 2011 18th International Conference
  on}{\nolinebreak\hspace{0.1em}]},   1--4, IEEE (2011).

\bibitem{dempster1977maximum}
Dempster, A., Laird, N., and Rubin, D., ``Maximum likelihood from incomplete
  data via the em algorithm,'' {\em Journal of the Royal Statistical Society.
  Series B (Methodological)} ,  1--38 (1977).

\bibitem{o2007theoretical}
O'Toole, A., Jiang, F., Abdi, H., P{\'e}nard, N., Dunlop, J., and Parent, M.,
  ``Theoretical, statistical, and practical perspectives on pattern-based
  classification approaches to the analysis of functional neuroimaging data,''
  {\em Journal of cognitive neuroscience}~{\bf 19}(11),  1735--1752 (2007).

\bibitem{yang2003can}
Yang, J. and Yang, J., ``Why can lda be performed in pca transformed space?,''
  {\em Pattern recognition}~{\bf 36}(2),  563--566 (2003).

\bibitem{sima2008peaking}
Sima, C. and Dougherty, E., ``The peaking phenomenon in the presence of
  feature-selection,'' {\em Pattern Recognition Letters}~{\bf 29}(11),
  1667--1674 (2008).

\bibitem{martinez2001pca}
Martinez, A. and Kak, A., ``Pca versus lda,'' {\em Pattern Analysis and Machine
  Intelligence, IEEE Transactions on}~{\bf 23}(2),  228--233 (2001).

\bibitem{ellis@mfcc}
Ellis, D., ``Reproducing the feature outputs of common programs using matlab
  and melfcc.m,'' (May 2005).
\newblock
  \texttt{http://http://labrosa.ee.columbia.edu/matlab/rastamat/mfccs.html}.

\bibitem{jurafsky2008speech}
Jurafsky, D. and Martin, J., ``Speech and language processing: An introduction
  to speech recognition,'' {\em Computational Linguistics and Natural Language
  Processing. 2nd Edn., Prentice Hall, ISBN}~{\bf 10}.

\end{thebibliography}
\bibliographystyle{spiebib}

\end{document}